\documentclass[12pt]{iopart}
\usepackage{iopams}  
\usepackage{graphicx}
\usepackage{cases}
\usepackage{url}

\usepackage{natbib}

\newcommand{\be}{\begin{equation}}
\newcommand{\ee}{\end{equation}}
\newcommand{\bea}{\begin{eqnarray}}
\newcommand{\eea}{\end{eqnarray}}
  
\newcommand{\pp}{\varphi}  

\newcommand{\dd}{\mathrm{d}}

\begin{document}

\title{Testing Chern-Simons gravity with black holes?}

\author{F H Vincent$^{1}$}
\address{$^{1}$ Nicolaus Copernicus Astronomical Center, ul. Bartycka 18, PL-00-716 Warszawa, Poland}
\ead{fvincent@camk.edu.pl}

\begin{abstract}

We investigate the possibility to distinguish the small-coupling, slow-rotation
black hole solution of Chern-Simons (CS) gravity from the Kerr solution. We develop
simulations of electromagnetic observables in the vicinity of CS and Kerr black holes.
We show that the typical relative observable difference between CS and Kerr spacetimes is of
the order of $0.1\%$ thus beyond reach of current or near-future instruments. 

\end{abstract}



\section{Introduction}

General relativity (GR) has so far passed all tests with flying colors~\citep{will09}.
However, it remains to be more thoroughly tested in the strong-field regime.
Black holes (BH) are a perfect environment for such tests, being the most compact
astrophysical objects. In this article, we examine the possibility to distinguish GR
from an alternative theory of gravitation that attracted considerable attention recently,
Chern-Simons (CS) dynamical gravity~\citep{alexander09}.

We study here the observable properties of the \emph{slowly rotating} black hole solution derived recently
in Refs~\citep{yunes09,konno09}, in the limit of a \emph{small coupling} (see a precise definition below). 
We restrict ourselves to the \emph{electromagnetic signatures}
of both classes of BH, stellar-mass and supermassive ones. 
In this perspective, we will focus on studying the effect of CS gravity on:
\begin{itemize}
\item the Galactic center supermassive BH Sgr~A* silhouette,
\item the continuous X-ray flux of stellar-mass black holes in X-ray binaries,
\item the iron line profile of both classes of BH,
\item the quasi-periodic oscillations of both classes of BH.
\end{itemize}
Our aim is to present a quantitative analysis of the impact of CS gravity
on all these observables in order to determine whether current or near-future 
electromagnetic observations could allow distinguishing between GR and CS gravity. 

Few studies have been devoted so far to the electromagnetic signatures of CS gravity.
Two recent studies have been dedicated to the difference of flux emitted by a thin accretion
disk surrounding a Kerr or CS black hole~\citep{harko10} and to the impact of CS gravity
on the silhouette of a BH~\citep{amarilla10}. Our aim is to give a broader description of electromagnetic
signatures of CS slowly rotating BH and to formulate the results in a format that can be readily
compared to real observed data.


Section~\ref{sec:CS} briefly describes Chern-Simons dynamical gravity. Section~\ref{sec:CSBH}
presents the slowly rotating CS black-hole solution and geodesic motion in such a spacetime.
Section~\ref{sec:res} analyzes various electromagnetic signatures of CS slowly rotating BH and
section~\ref{sec:ccl} gives conclusions and perspectives.

\section{Dynamical Chern-Simons gravity}
\label{sec:CS}

Dynamical Chern-Simons gravity is described by the action

\be
S = S_{\mathrm{EH}} + S_{\mathrm{CS}}  + S_{\vartheta}  + S_{\mathrm{mat}}  
\ee
where the right-hand terms are defined below.
The action
\be
S_{\mathrm{EH}} = \kappa \int \dd x^4 \sqrt{-g} R
\ee
is the Einstein-Hilbert action, with $\kappa=1/16 \pi$ (here and in the following, we use units in which
$c=G=1$), $g$ being the metric determinant and $R$ the
Ricci scalar.
The CS correction is described by
\be
S_{\mathrm{CS}} = \frac{\alpha}{4} \int \dd x^4 \sqrt{-g} \,\vartheta  \; {}^* R R
\ee
with $\alpha$ being a coupling constant, $\vartheta$ the CS coupling scalar field describing deformation
from GR (a constant $\vartheta$ reduces CS gravity to GR),
and ${}^* R R = 1/2 \,\epsilon^{\alpha\beta\mu\nu} R_{\alpha\beta\gamma\delta}R^{\gamma\delta}_{\:\:\:\:\:\mu\nu}$ is the 
Pontryagin density defined by the contraction of the Riemann tensor and its dual, $\epsilon$ being the Levi-Civitta tensor.
The action
\be
S_{\vartheta} = -\frac{\beta}{2} \int \dd x^4 \sqrt{-g} \left[ g^{\mu\nu} \nabla_{\mu} \vartheta\,\nabla_{\nu} \vartheta + 2\,V\left(\vartheta\right)\right]
\ee
is the scalar field action, sum of a kinetic and potential terms, with $\beta$ being a coupling constant.
Finally
\be
S_{\mathrm{mat}} =  \int \dd x^4 \sqrt{-g} \mathcal{L}_{\mathrm{mat}}
\ee
is the matter action.

In the perspective of testing the theory, it is important to constrain the coupling constants. It is useful
to introduce two other coupling parameters:
\be
\xi \equiv \frac{\alpha^2}{\kappa\,\beta}
\ee
has the dimension of $L^4$, where $L$ is a unit of length. CS gravity reduces to GR in the limiting case $\xi=0$.
The dimensionless parameter
\be
\zeta \equiv \frac{\xi}{M^4},
\ee
which scales with the typical mass $M$ of the system, will appear in the expression of the metric for slowly rotating
CS black holes. The slowly rotating BH solution found in Refs.~\citep{yunes09,konno09} is only valid in the limit
of small coupling : $\zeta \ll 1$.

To date the best constraint on the coupling constant of the theory is given by~\citep{haimoud11}

\be
\xi^{1/4} \lesssim 10^8 \,\mathrm{km}.
\ee
Note that a much more stringent limit was provided by Ref.~\citep{yunes09} but assuming that the
exterior metric of a CS neutron star can be described by the slowly rotating BH metric, which is
not the case as demonstrated in Ref.~\citep{haimoud11}. 
The corresponding dimensionless coupling parameter is restricted to $\zeta \lesssim 10^7$ for
a $10^6 M_{\odot}$ BH, and to $\zeta \lesssim 10^{27}$ for a $10 M_{\odot}$ BH. The only coupling limit
that we will consider in this article is thus the small-coupling condition $\zeta \ll 1$.


\section{Geodesic motion around Chern-Simons slowly rotating black holes}
\label{sec:CSBH}

\subsection{Slowly rotating CS black holes}

The metric of a slowly rotating CS black hole is given by~\citep{yunes09,konno09}

\bea
\label{eq:met}
\dd s^2 = &-&\left(f+\frac{2 a^2}{r^3}\mathrm{cos}^2\theta\right)\dd t^2 \\ \nonumber
	       &+&\left(\frac{1}{f} - \frac{a^2}{f r^2}\left[\frac{1}{f} - \mathrm{cos}^2 \theta \right]\right)\dd r^2 \\ \nonumber
	       &+&\left(r^2 + a^2\mathrm{cos}^2 \theta \right)\dd \theta^2 \\ \nonumber
	       &+&\left(r^2 + a^2\left[1 + \frac{2}{r}\mathrm{sin}^2 \theta \right]\right)\dd \pp^2 \\ \nonumber
	       &+&\left(-\frac{4a}{r}\mathrm{sin}^2\theta + \frac{5 a \zeta}{8 r^4}\left[1+\frac{12}{7r}+\frac{27}{10 r^2}\right]\mathrm{sin}^2 \theta\right)\dd t \dd \pp \\ \nonumber
\eea
where $f=1-2/r$, and we have chosen units in which the black hole mass $M$ is unity. 
Here the metric signature is $(-,+,+,+)$. Note that this metric reduces to the slow-rotation limit of the Kerr metric
when $\zeta=0$, and that the above solution is only correct in the limit $\zeta \ll 1$, $a \ll 1$. Only the $g_{t\pp}$ term of the metric is 
modified, and it is very clear that only strong-field phenomena will allow distinguishing this solution from the Kerr solution
as the CS correction to $g_{t\pp}$ is strongly damped by a $1/r^4$ factor.

\subsection{Equation of geodesics}

Let us consider a particle with mass $\mu$ in the spacetime described above. Its specific
energy $E=-u_t$ and specific angular momentum $L=u_\pp$ are conserved in geodesic motion.
The geodesic equation reads~\citep{sopuerta09}

\bea
\label{eq:geo}
\Sigma \,\dot{t} &=& \left[ -a \left(a E \mathrm{sin}^2 \theta - L\right)+\left(r^2+a^2\right)\frac{P}{\Delta}\right] + \Sigma L \delta_{\mathrm{CS}}, \\ \nonumber
\Sigma^2 \dot{r}^2 &=& \left[P^2 - \Delta \left\{\mu^2 r^2 + \left(L-aE\right)^2 +Q \right\} \right] + 2ELf \Sigma^2 \delta_{\mathrm{CS}}, \\ \nonumber
\Sigma^2 \dot{\theta}^2 &=& \left[Q - \mathrm{cos}^2 \theta \left\{ a^2 \left(\mu^2 - E^2\right) + \frac{L^2}{\mathrm{sin}^2 \theta} \right\} \right], \\ \nonumber
\Sigma \,\dot{\pp} &=& \left[ -\left(aE-\frac{L}{\mathrm{sin}^2 \theta}\right) + \frac{a}{\Delta} P\right] - \Sigma E \delta_{\mathrm{CS}}. \\ \nonumber
\eea
where $\Sigma = r^2 + a^2 \mathrm{cos}^2\theta$, $\Delta = r^2 - 2r +a^2$, $P = E\left(r^2+a^2\right) - aL $ and $Q$ is 
the Carter constant, that is still conserved in CS slowly rotating black hole solution with the same expression as in Kerr.
The CS correction term reads~\citep{sopuerta09}
\be
\delta_{\mathrm{CS}} = \frac{a\zeta}{112\, r^8 f} \left(70\, r^2 +120 \,r +189\right).
\ee

The quantities appearing in brackets on the right-hand sides of equation~(\ref{eq:geo}) are the standard Kerr geodesic equation. To ease the
interpretation of the above equations, these Kerr terms were not approximated by their slow-rotation expression. This
approximation was done to obtain the results presented in Section~\ref{sec:res}.

\subsection{Quasi-circular motion}

The results presented in Section~\ref{sec:res} imply computing some standard quantities of quasi-circular motion
in an axially-symmetric stationary spacetime. For completeness the expressions of these quantities are given here.
Most of them are taken from Refs.~\citep{bardeen72,harko10}.

\paragraph{4-velocity of circular motion}
The 4-velocity of a massive particle orbiting around a BH in exactly circular motion, is

\be
\mathbf{u} = u^t\left(1,0,0,\Omega\right)
\ee 
where
\bea
\Omega &=& \frac{-g_{t\pp,r}+\sqrt{\left(g_{t\pp,r}\right)^2-g_{tt,r}g_{\pp\pp,r}}}{g_{\pp\pp,r}} \\ \nonumber
u^t &=& \frac{1}{\sqrt{-g_{tt}-2g_{t\pp}\Omega-g_{\pp\pp}\Omega^2}} \\ \nonumber
\eea

\paragraph{Innermost stable circular orbit}
The radius of the innermost stable circular orbit (ISCO) is found by imposing that the effective
potential and its first and second derivatives are zero. The ISCO radius is thus found to be the non-zero real root
of the following equation:

\be
E^2 g_{\pp\pp,rr} + 2ELg_{t\pp,rr}+L^2 g_{tt,rr} - \left(g_{t\pp}^2 - g_{tt}g_{\pp\pp}\right)_{,rr} = 0
\ee
where the constant of motion can be related to the metric coefficients:

\bea
\label{eq:EL}
E &=& -\frac{g_{tt}+g_{t\pp}\Omega}{\sqrt{-g_{tt}-2g_{t\pp}\Omega-g_{\pp\pp}\Omega^2}}, \\ \nonumber
L &=& \frac{g_{t\pp}+g_{\pp\pp}\Omega}{\sqrt{-g_{tt}-2g_{t\pp}\Omega-g_{\pp\pp}\Omega^2}}. \\ \nonumber
\eea

\paragraph{Photon orbit}
Not all values of $E$ and $L$ are allowed for a particle moving on a circular orbit around a BH. In
order for equations~(\ref{eq:EL}) to be valid, the argument of the square root must be positive. The
limiting case of zero denominator corresponds to the circular orbit of a particle with infinite specific energy,
thus to a photon orbit. It is the innermost limit of all circular orbits of particles. The radius of the photon orbit
is thus the real non-zero root of:

\be
g_{tt}+2g_{t\pp}\Omega+g_{\pp\pp}\Omega^2=0.
\ee

\paragraph{Flux of a geometrically thin disk}
Ref.~\citep{page73} gives the expression of the flux emitted by a geometrically thin accretion
disk in circular rotation:

\be
F(r) = \frac{\dot{M}}{4\pi} \frac{1}{\sqrt{g_{rr}\left(g_{t\pp}^2 - g_{tt}g_{\pp\pp}\right)}} \frac{-\Omega_{,r}}{\left(E-\Omega L\right)^2} \int_{r_{\mathrm{ms}}}^{r} \left(E - \Omega L\right) L_{,r} \mathrm{d}r
\label{eq:flux}
\ee
where $\dot{M}$ is the averaged accretion rate and $r_{\mathrm{ms}}$ is the ISCO radius (assuming
the inner edge of the disk is at ISCO).

\paragraph{Epicyclic pulsations}
Let us now consider a massive particle in quasi-circular motion. It will oscillate with radial
and vertical epicyclic frequencies given by (see~\citep{abramowicz03})

\bea
\omega_r^2 &=& \frac{\left(g_{tt}+\Omega g_{t\pp}\right)^2}{2g_{rr}} \left. \frac{\partial^2 \mathcal{U}}{\partial r^2}\right|_{l} \\ \nonumber
\omega_\theta^2 &=& \frac{\left(g_{tt}+\Omega g_{t\pp}\right)^2}{2g_{\theta\theta}} \left. \frac{\partial^2 \mathcal{U}}{\partial \theta^2}\right|_{l} \\ \nonumber
\eea
where the effective potential $\mathcal{U}$ is given by (see e.g.~\citep{abramowicz09})

\be
\mathcal{U} = g^{tt} - 2 l g^{t\pp} + l^2 g^{\pp\pp}
\ee
and $l = L/E$.

All the quantities given above can be computed straightforwardly provided the metric is known. In the next section, 
they will be computed using the metric~(\ref{eq:met}).

\section{Strong-field electromagnetic signatures of Chern-Simons gravity}
\label{sec:res}

This section is devoted to determine quantitatively the difference between a CS
black hole as described by metric~(\ref{eq:met}) and a Kerr BH of identical spin,
for various observables. 

In order to abide by the two conditions of slow rotation and small coupling, \emph{we
fix the spin and dimensionless coupling parameter to $(a,\zeta) = (0.1,0.1)$}. These
two typical values will allow us to derive \emph{characteristic values of the observable difference between CS
slow-rotation small-coupling solution and the Kerr metric}.

All the ray-traced simulations presented in this section were computed using the open-source\footnote{Freely accessible at \url{gyoto.obspm.fr}.}
code \texttt{GYOTO}~\citep{vincent11}.

\subsection{Sgr~A* silhouette}

Here, we compute the silhouette of the Galactic center black hole Sgr~A*, assuming it is 
surrounded by a geometrically thin, optically thick accretion disk extending from the
ISCO to $r=20$. The flux emitted by such a disk is derived from equation~(\ref{eq:flux}).
We are interested
in determining the difference of angular size of the black hole silhouette between Kerr
and CS spacetimes. Let us recall that the angular size of the silhouette of a black hole
is \emph{only depending on gravitational effects}, not on the astrophysical assumptions we
made on the disk structure and emission (including the assumption that the disk terminates
at ISCO). Measuring the size of a black hole silhouette is thus a very powerful, astrophysically-unpolluted
probe of spacetime.

Photons contributing to the silhouette of the black hole come at closest approach at a coordinate
radius equal to the photon orbit radius. For our choice of parameters, the corotating Kerr and CS photon
orbits are located at

\be
r_{\mathrm{ph,Kerr}}=2.88219, \quad r_{\mathrm{ph,CS}}=2.88287,
\ee
thus differing by $0.02\%$. It is very clear from this number only that no difference whatsoever
will be observable on the black hole image.

\begin{figure}
	\centering
	\includegraphics[height=5cm,width=5cm]{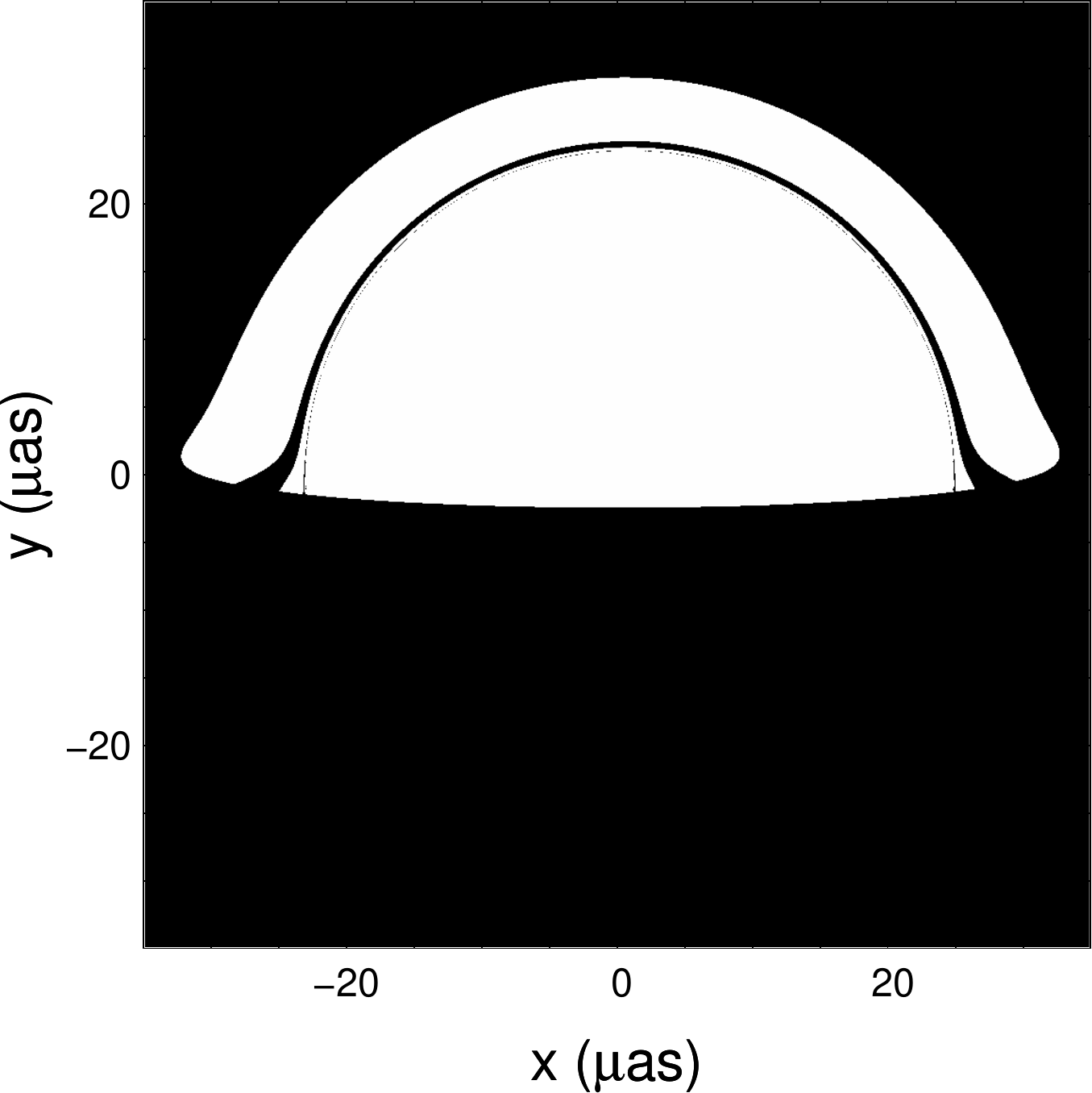} \hspace{0.5cm}
	\centering
	\includegraphics[height=7cm,width=5cm]{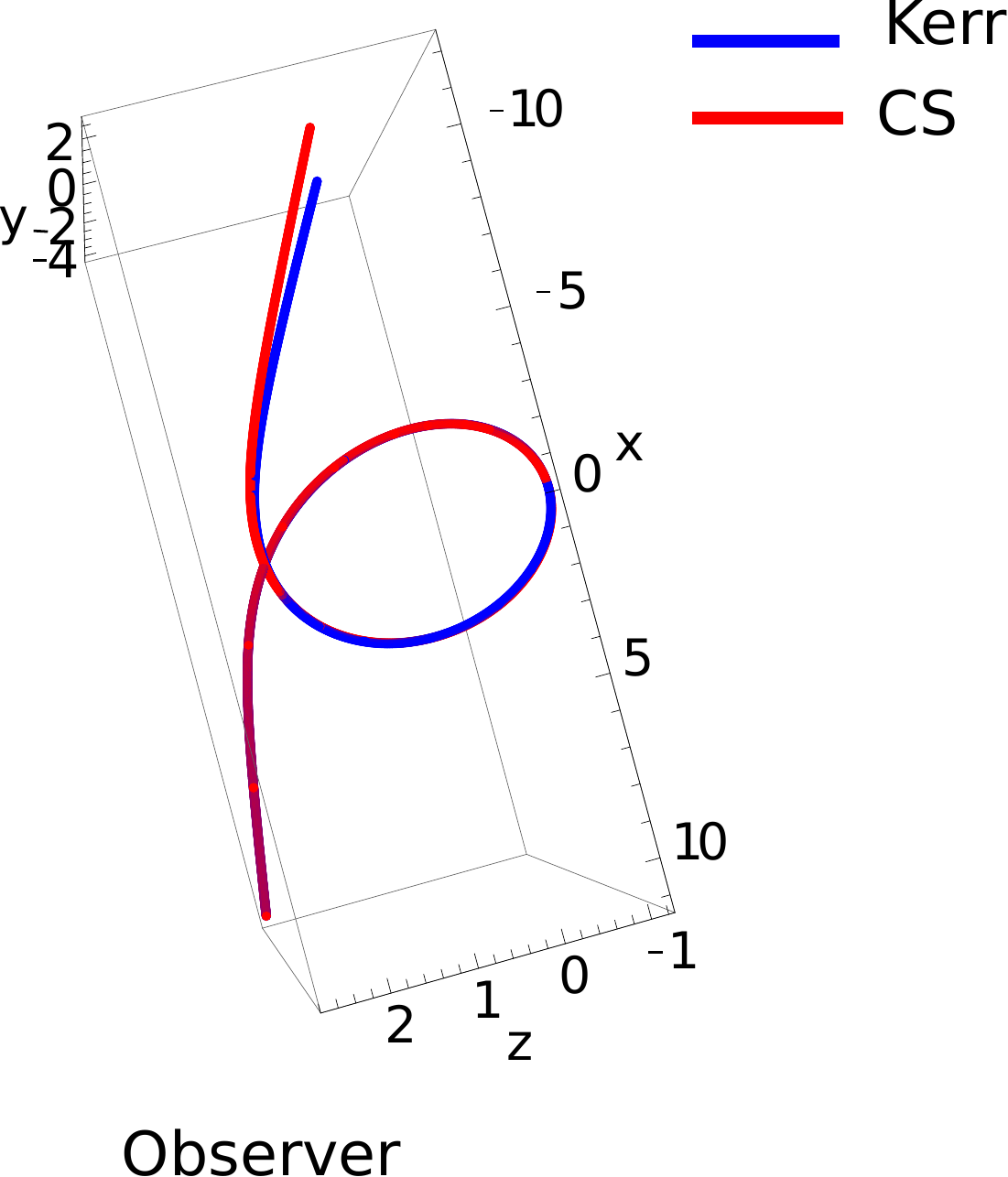}
	\caption{{\bf Left:} superimposition of the images of the innermost part of a geometrically thin optically thick accretion disk around a CS and Kerr black hole at the Galactic center
		as observed from Earth, with $(a,\zeta)=(0.1,0.1)$ and an inclination of $85^{\circ}$. 
		The axes are graduated in $\mu$as. Illuminated pixels are shown in black whereas non-illuminated
		pixels are in white color.
		The silhouette is the
		thin half-ring of illuminated pixels at the center of the image. The two silhouettes angular sizes are indistinguishable. {\bf Right:} null geodesics of a photon contributing 
		to the silhouette in Kerr (blue) and CS (red) spacetimes. The observer is located downwards. The axes are graduated in natural units ($G=M=c=1$).}
	\label{fig:silh}
\end{figure}

Figure~\ref{fig:silh} shows the superimposed images of CS and Kerr black holes, together with the null geodesic of
a photon contributing to the silhouette. The images of the
Kerr and CS black holes are undistinguishable by eye at the resolution of the figure ($1000\times1000$
pixels). The difference between the two images is of order of the pixel size,
which is approximately $0.01\,\mu$as, thus far below the resolution of any instrument even
of the forseeable future\footnote{The Event Horizon Telescope~\citep{doeleman09}
will allow imaging the silhouette of Sgr~A* with the resolution of 1~$\mu$as.}. 
The right panel shows that the trajectories of photons that come very close to the black
hole, when integrated backward in time from the observer to the disk, are clearly different. 
However their apparent directions on the observer's sky are the same. Only
the intensity transported by these photons will differ, but its value depends on astrophysical assumptions
and are not as clear a probe as the angular size of the silhouette.

\subsection{X-ray flux from stellar-mass black holes}

The impact of CS gravity on the X-ray flux emitted by black-hole X-ray binaries
is important as these objects are used in order to derive constraints on the Kerr spin
parameter through the continuum-fitting technique (for a review, see Ref.~\citep{mcclintock11}),
assuming GR is the correct description of gravitation.

Here we consider an optically thick, geometrically thin disk, the inner radius of which coincides
with the ISCO, and with outer radius $20 \,M$. The ISCO radius in the Kerr and CS cases with
our choice of parameters is given by
\be
r_{\mathrm{ISCO,Kerr}}=5.6693, \quad r_{\mathrm{ISCO,CS}}=5.6698,
\ee
thus differing by $0.01\%$ (with the CS ISCO greater than the Kerr ISCO).
The flux emitted by such a disk is given by equation~(\ref{eq:flux}).
In order to retrieve a result comparable to observed data, this equation must be multiplied by a factor $c^6/G^2 M^2$.
Two parameters must then be chosen, the mass $M$ and accretion rate $\dot{M}$ of the source. Moreover,
the source distance $D$ will also appear to get observed flux values. We consider here a typical black-hole
X-ray binary with $M=10\,M_{\odot}$, $\dot{M}=10^{18}\, \mathrm{g\,s^{-1}}$ and $D=10\,\mathrm{kpc}$.

The emitted flux $F_{\mathrm{em}}$ being known at any point of the disk, the corresponding temperature $T$ is simply given 
by Stefan-Boltzmann law $F_{\mathrm{em}} = \sigma T^{4}$ where $\sigma$ is Stefan-Boltzmann constant. Assuming
a blackbody radiation, the spectrum emitted by the disk is then at hand.
A practical analytical expression of equation~(\ref{eq:flux}) is given in Ref.~\citep{page73} in the Kerr case, which is
thus immediate to compute. In the CS case, we have computed a table of temperature values for a set of radii,
and interpolate to get the actual temperature at any point of the disk. 
Once a map of $B_{\nu}(T)$ values is computed, where $B_{\nu}$ is the Planck function, the observed flux
is computed according to the expression:

\be
F_{\nu,\mathrm{obs}} = \sum_{\mathrm{pixels}} B_{\nu}(T_{\mathrm{pixel}}) \,\mathrm{cos}\theta\, \frac{\Delta \Omega}{N_{\mathrm{pixels}}}
\ee
where the sum is performed over the \texttt{GYOTO} screen containing $N_{\mathrm{pixels}}$ pixels, $\theta$ is the angle between the normal
to the screen and the current pixel direction, and $\Delta \Omega = \pi L^2/D^2$ is the solid angle subtended by the screen
which is linked to the field-of-view size $L$ and to the distance of the source $D$.

\begin{figure}
	\centering
	\includegraphics[height=6.5cm,width=6.5cm]{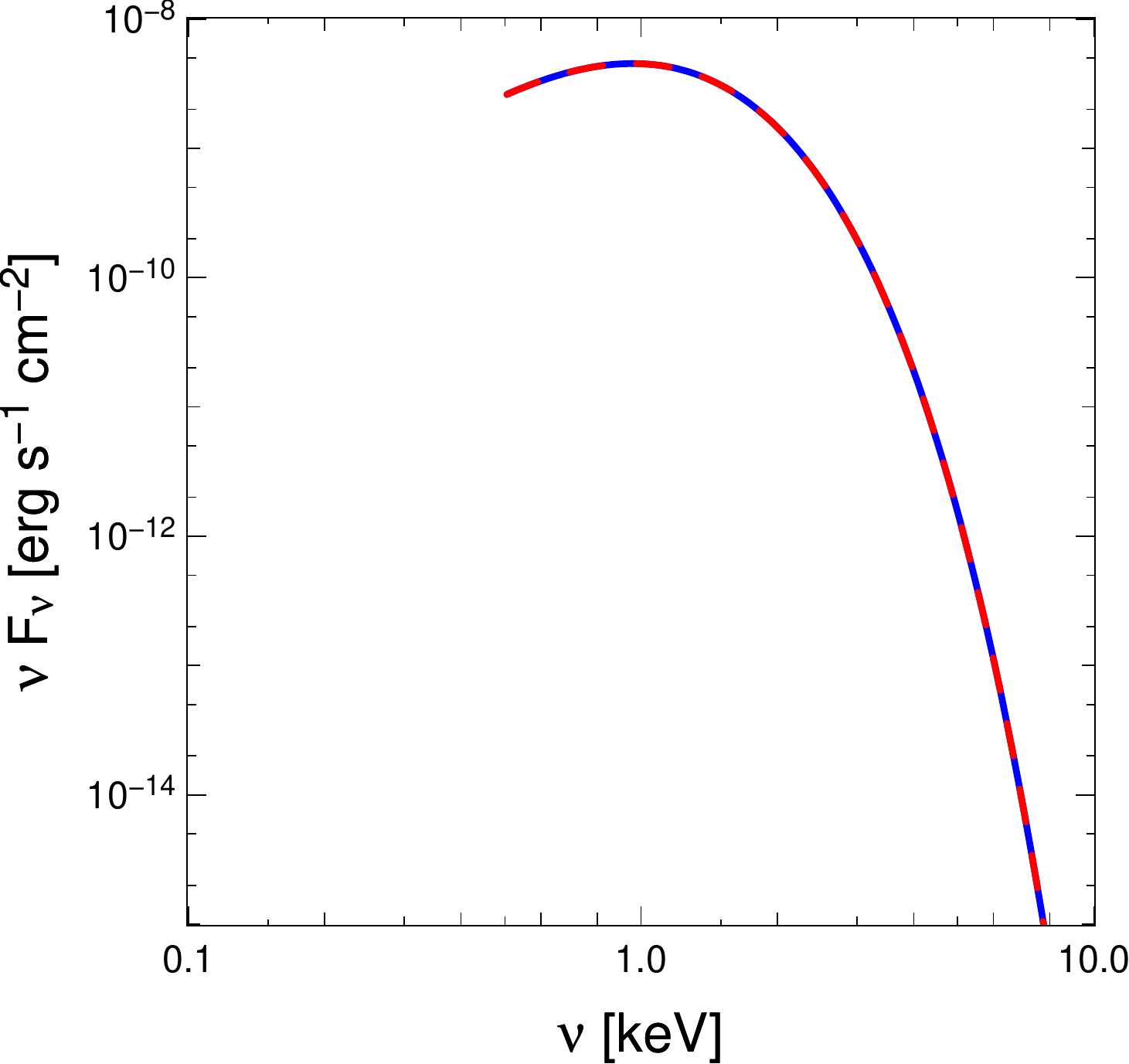}
	\includegraphics[height=6.5cm,width=6.5cm]{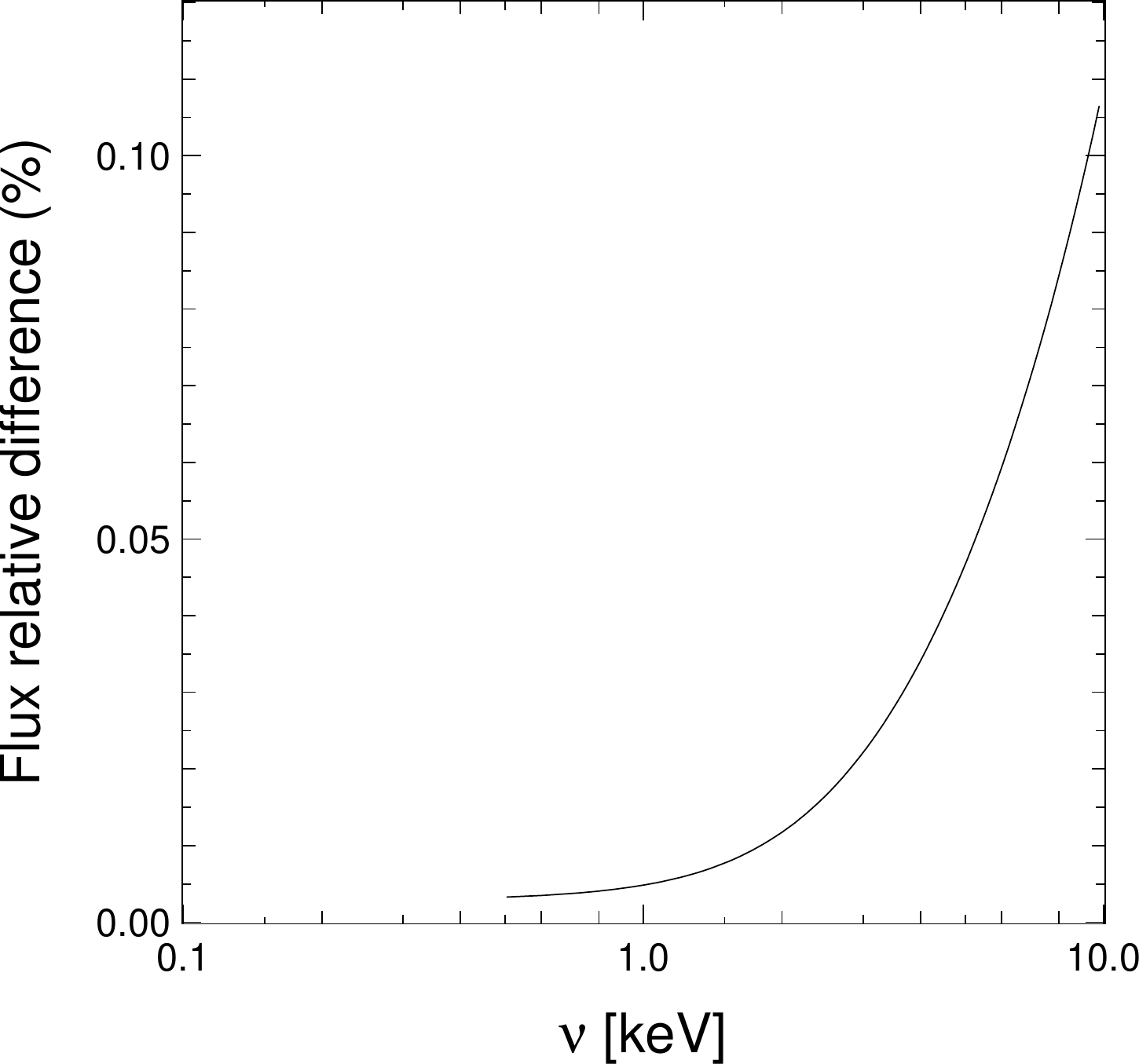}
	\caption{{\bf Left:} X-ray observed spectrum from a Page-Thorne accretion disk surrounding a Kerr (solid blue) or CS (dashed red)
		black hole with $(a,\zeta)=(0.1,0.1)$ and an inclination of $45^{\circ}$. 
		{\bf Right:} relative difference between the Kerr and CS fluxes.}
	\label{fig:spec}
\end{figure}

Figure~\ref{fig:spec} shows the spectrum, as observed from Earth, of a typical black-hole X-ray binary
in the Kerr and CS cases. The order of magnitude of the flux computed agrees well with real observed
data (see e.g. Fig.~6 of Ref.~\citep{mcclintock11}) The right panel shows that the relative difference is of typically $0.05\%$
which makes it impossible to distinguish with current instruments. The increase of the relative difference with frequency
is a consequence of the fact that the change of $B_{\nu}(T)$ with $T$ increases with higher frequencies, for the
range of temperatures considered here (i.e. a few $10^6\,\mathrm{K}$). It appears thus that a difference between Kerr
and CS thin accretion disks is easier at higher frequency. However, the difference, even at high frequency, stays too low
to be observable.

\subsection{Iron line profile}

Fitting of iron-line profiles is the other commonly used method of fitting spins of
black holes (see Ref.~\citep{reynolds03} for a review). It is thus also important
to determine the impact of alternative theories of gravitation on this observable.

Following Ref.~\citep{reynolds03}, we model iron lines spectra by considering an
optically thick, geometrically thin disk surrounding a black hole extending from the
ISCO to an outer radius of $r_{\mathrm{outer}}=50\,M$. The emitted specific intensity
is given by
\be
I_{\nu,\mathrm{em}} = \delta(\nu-\nu_{0}) \,r^{-\beta}
\ee
where $\beta=3$, $\nu_0=6.4\,\mathrm{keV}$, and $\delta$ is the Dirac distribution.
For numerical implementation, this distribution is approximated by

\begin{equation}
\delta(\nu-\nu_0)=
\left\lbrace
\begin{array}{ccc}
1  & \mbox{if} & \left|\nu-\nu_0\right|<0.01\,\nu_0\\
0 & \mbox{else} & \\
\end{array}\right.
\end{equation}
This choice is linked to the spectral resolution of CHANDRA which is of the order
of $E / \Delta E \approx 100$.

\begin{figure}
	\centering
	\includegraphics[height=6cm,width=6cm]{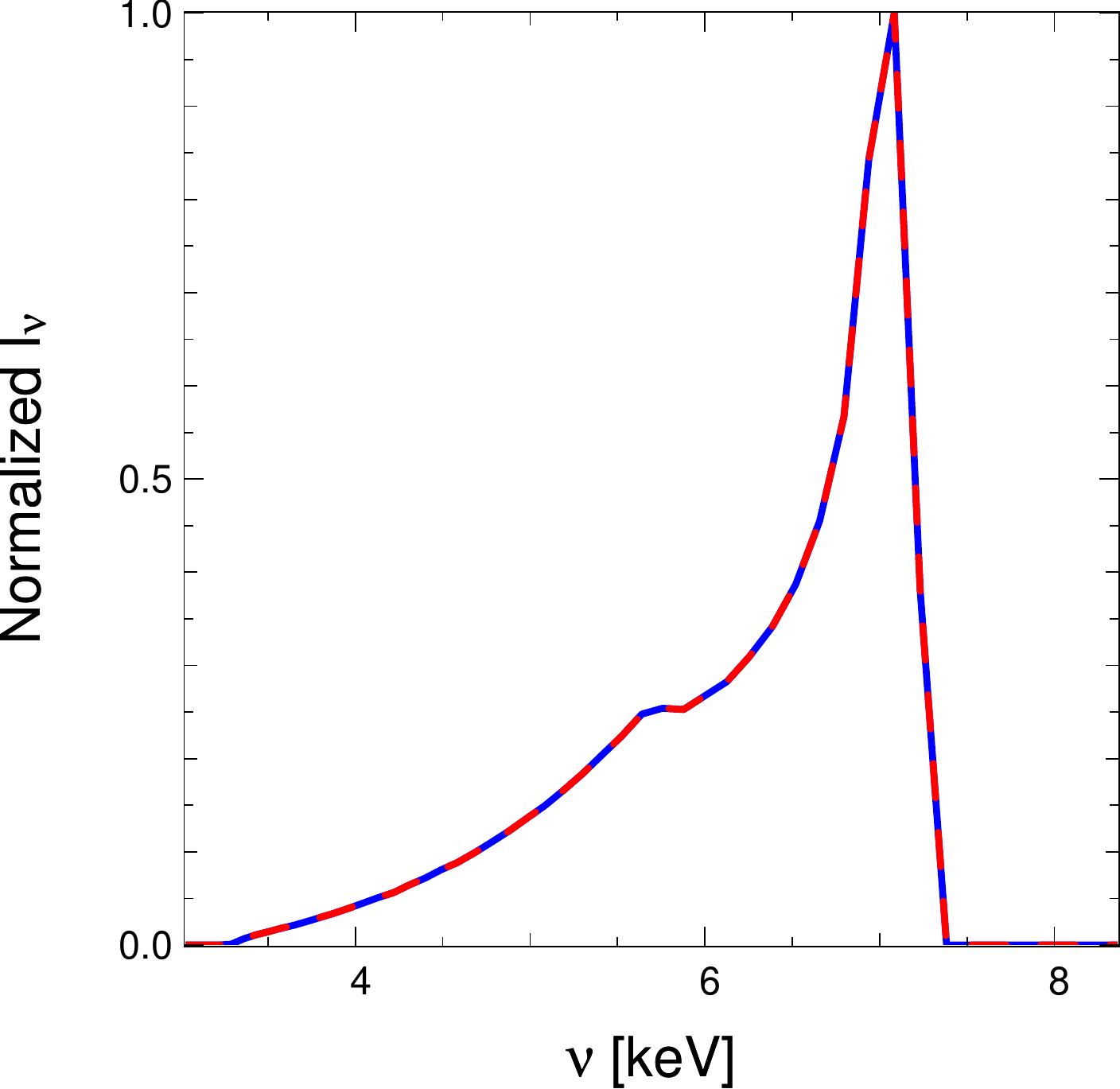}
	\includegraphics[height=6cm,width=6cm]{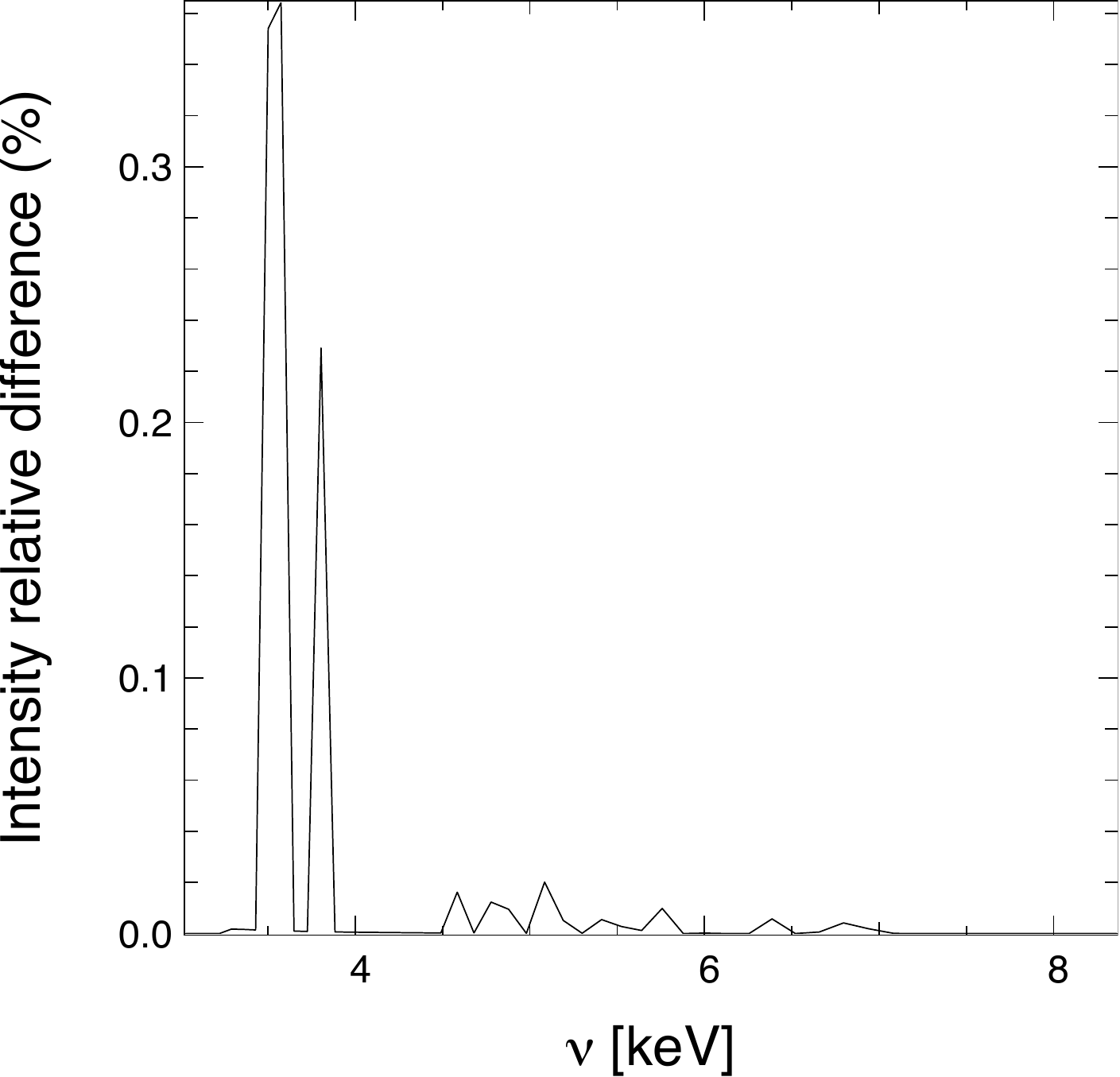}
	\caption{{\bf Left:} Iron line profile emitted by a geometrically thin disk surrounding a Kerr (solid blue) or CS (dashed red)
		black hole with $(a,\zeta)=(0.1,0.1)$ and an inclination of $45^{\circ}$. {\bf Right:} 
		relative difference between the Kerr and CS profiles.}
	\label{fig:iron}
\end{figure}

Figure~\ref{fig:iron} shows the corresponding ray-traced iron line profile of the observed specific intensity $I_{\nu,\mathrm{obs}}$,
where only the contribution of the primary image was conserved. 
The overall aspect
can be compared to figure~11 of Ref.~\citep{reynolds03}.
The relative difference between the Kerr and CS profiles is of order $0.1\,\%$, thus below
detection limit. 
Moreover, the two peaks of relative difference that are the only ones reaching the $0.1\,\%$ level
are due to the limited screen resolution. The height of these peaks is a decreasing function of the number
of screen pixels (figure~\ref{fig:iron} was obtained using $1500\times1500$ pixels images). This effect is due to the fact that
the emitting zone of the disk (i.e. the zone where the emitted frequency is close enough to the
line frequency) can be extremely thin projected on the observer's sky for some values of the
observed frequency. Then, high resolution
is important to resolve these thin areas. Going to even higher resolution would decrease the height
of these peaks, but as we are only interested here in determining whether the difference between Kerr
and CS gravity is observable, we do not care about the actual height of these peaks.

Let us note finally that given the emitted specific intensity is the same on both cases
here, the difference between the two profiles is due to the difference of photon geodesic motion
close to the black hole.

\subsection{Quasi-periodic oscillations}

A few microquasars exhibit high-frequency quasi-periodic oscillations (QPOs) that most probably are
linked to strong-field phenomena~(for a review, see Refs~\citep{remillard06}). The Galactic center flares
are also interpreted by some authors as QPOs~(see Refs~\citep{genzel10,morris12} for a review). Many models
have been proposed to account for such oscillations, and it is not the aim of this article to
review all of them in CS gravity. We will rather restrict ourselves here to two models, namely the
epicyclic resonance model~\citep{kluzniak01,abramowicz01} and the hot spot model~\citep{schnittman04}.

\begin{figure}
	\centering
	\includegraphics[height=5cm,width=7.5cm]{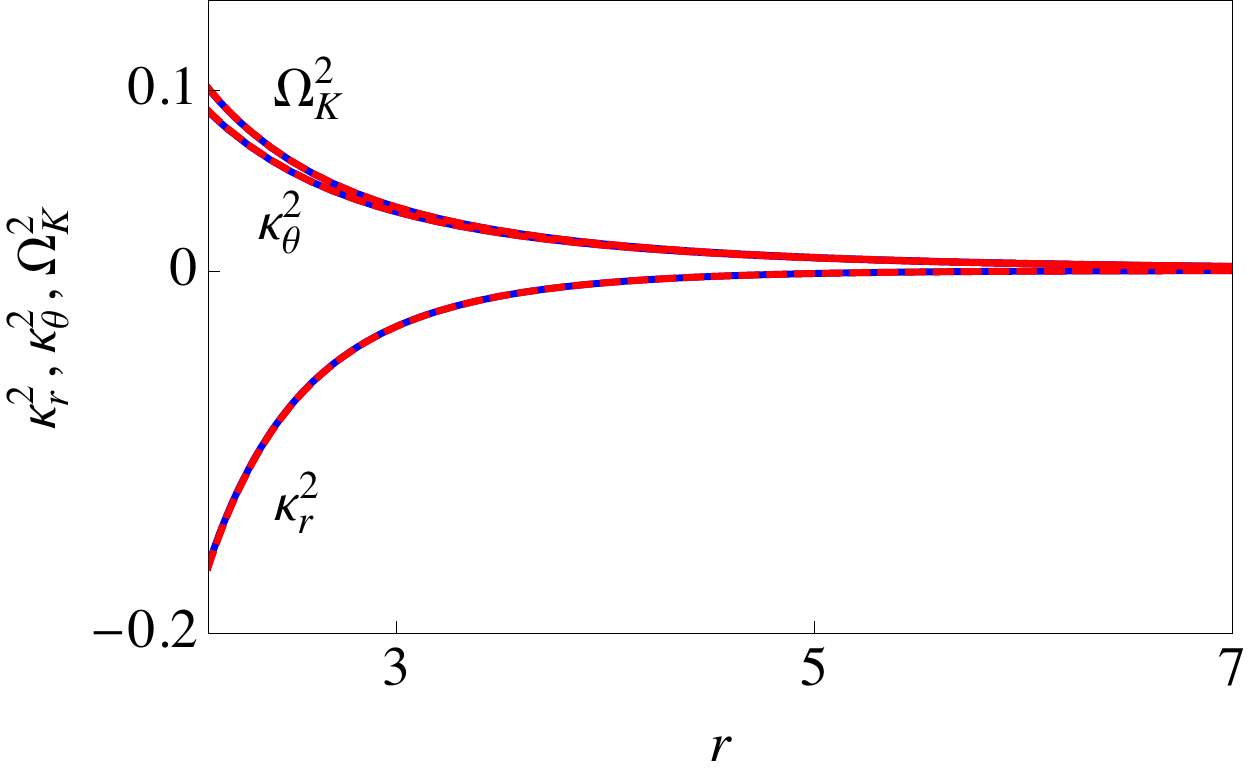}
	\includegraphics[height=5cm,width=7.5cm]{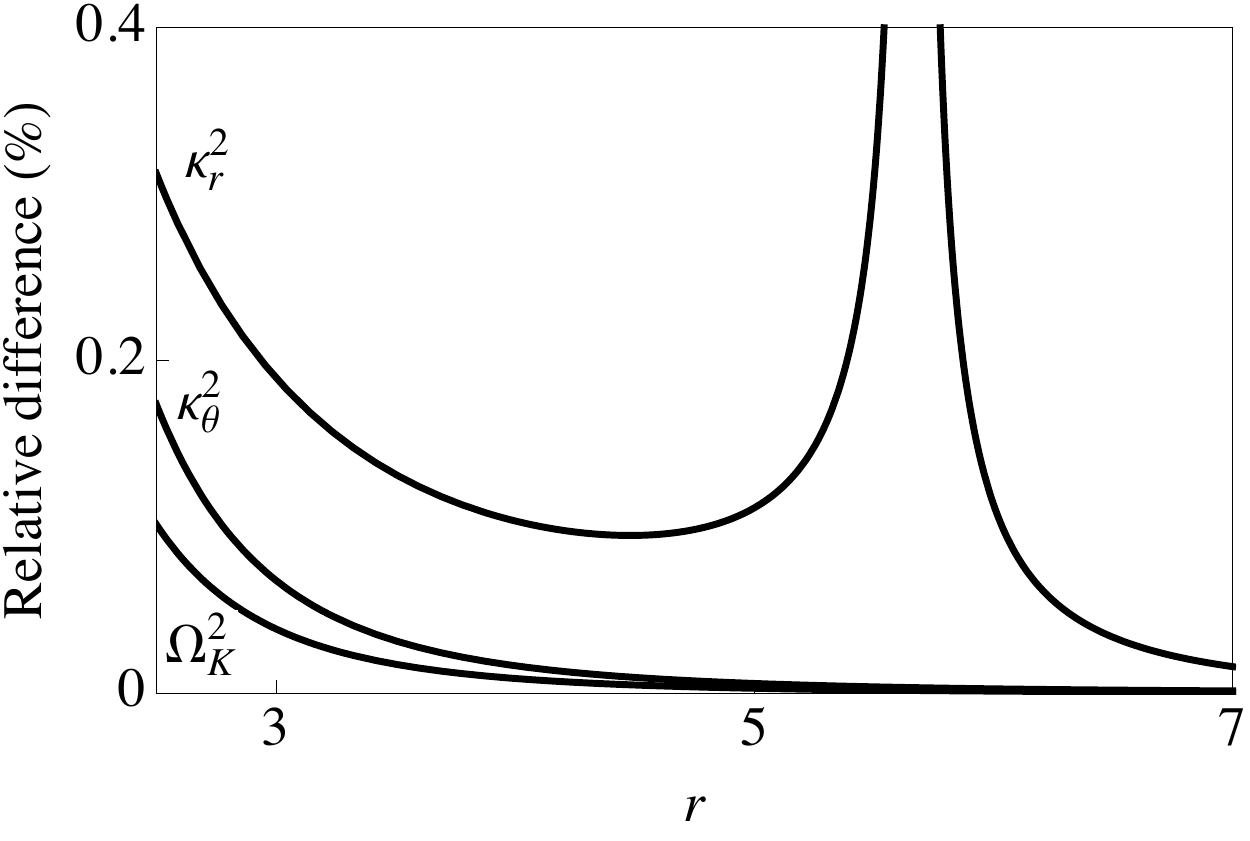}
	\caption{{\bf Left:} Keplerian ($\Omega_K$) and epicyclic radial ($\kappa_r$) and vertical ($\kappa_\theta$)
		pulsations profiles for $(a,\zeta)=(0.1,0.1)$ for Kerr (solid blue) and CS (dashed red) black hole. Natural
		units $G=M=c=1$ are used here. {\bf Right:} 
		relative difference between the Kerr and CS profiles. The radial epicyclic pulsation relative difference diverges at ISCO as
		both Kerr and CS pulsations are zero at this location, and the Kerr and CS ISCO are extremely close.}
	\label{fig:epifreq}
\end{figure}

\paragraph{Epicyclic resonance}
The epicyclic resonance model suggests that QPOs are due to a resonance between
Keplerian and epicyclic frequencies of a particle orbiting in a quasi-circular orbit
around a BH. It is thus interesting to determine the relative difference between Kerr and
CS epicyclic frequencies: this is shown in figure~\ref{fig:epifreq}. 
This figure shows that the
relative difference, for the typical spin and coupling parameter chosen in this section,
is of order of $0.1\%$ in the very neighborhood of the horizon, decreasing very
rapidly when getting further. The difference between epicyclic pulsations is thus below detection level.

\paragraph{Hot spot}
We have modeled a hot spot orbiting around a black hole following the work presented in Ref.~\citep{schnittman04}.
We consider a timelike geodesic orbiting circularly around a Kerr black hole of spin $0.1$ at coordinate radius $r=7$.
The hot spot is defined by a radius $R_{\mathrm{spot}}=1$. It is assumed to follow the guiding timelike geodesic and to emit radiation 
isotropically with gaussian modulation of full width at half maximum $\sigma=R_{\mathrm{spot}}/4$. The evolution of such a hot spot is 
then ray-traced both in Kerr and CS spacetimes, assuming its orbit is inclined at $45^{\circ}$ relative to the
distant observer. The only difference that affects the simulated light curve is thus
the difference of geodesic motion between the two spacetimes. Our interest here is to determine to what level the
difference of geodesic motion between Kerr and CS will impact the hot spot observable.
\begin{figure}
	\centering
	\includegraphics[height=6cm,width=7.5cm]{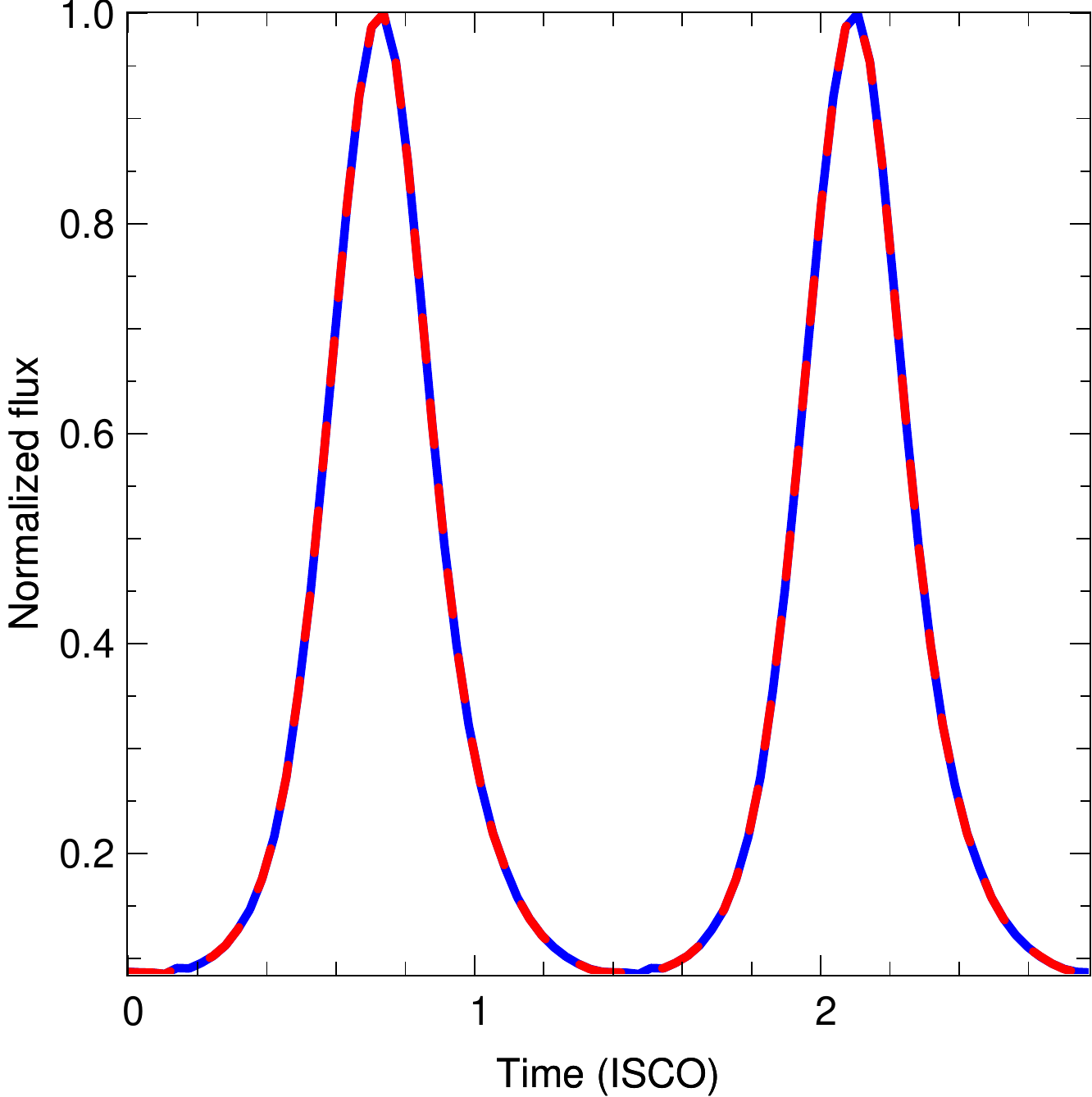}
	\includegraphics[height=6cm,width=7.5cm]{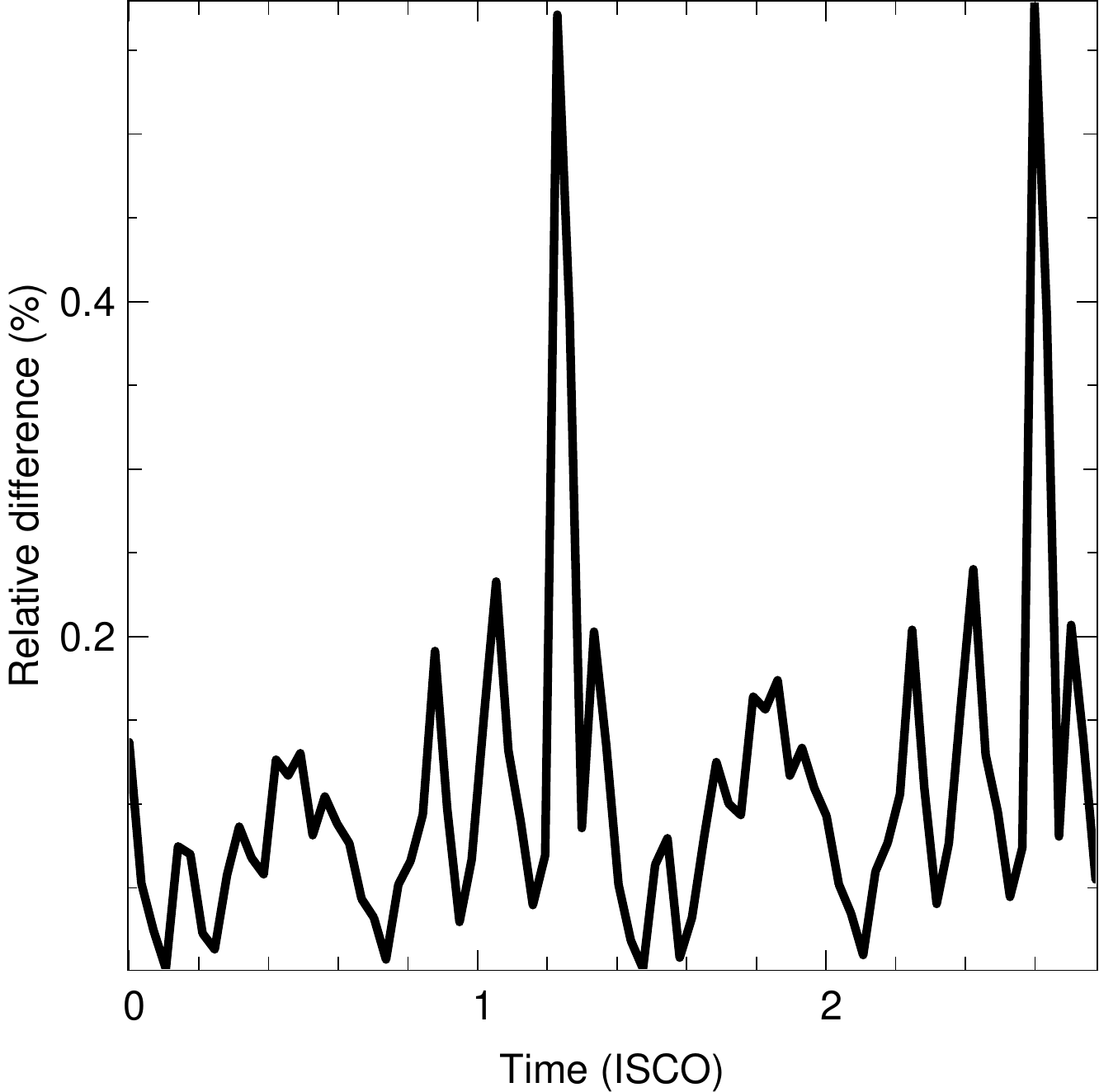}
	\caption{{\bf Left:} Hot spot light curve for Kerr (solid blue) and CS (dashed red) black hole, for $(a,\zeta)=(0.1,0.1)$ and
	an inclination of $45^{\circ}$.  {\bf Right:} 
		relative difference between the Kerr and CS light curves. The time unit is the ISCO period for Kerr spacetime with spin $0.1$.}
	\label{fig:hotspot}
\end{figure}
Figure~\ref{fig:hotspot} shows the hot spot's light curve for Kerr and CS spacetimes, together with their
relative difference. Here as well, the relative difference is of order a few $0.1\%$ which makes it impossible
to distinguish with present-day instrumental precision. Moreover, the same remark applies here as for the
iron line relative error in figure~\ref{fig:iron}: the two peaks of error that reach the $0.5\%$ level are due to
the contribution of the third-order images of the hot spot, which at the resolution used ($1000\times1000$ pixels)
are very thin. Thus the height of these peaks depends on resolution, but here again we are only interested
in demonstrating that the difference between Kerr and CS spacetimes is beyond instruments' reach.


\section{Conclusion}
\label{sec:ccl}

We have analyzed observables currently proposed as probes of strong-field general relativity
in the environment of black holes. Silhouettes, X-ray continuum flux, iron line profile, epicyclic pulsations
and hot spot evolution were simulated, both in Kerr and in slow-rotation, small-coupling Chern-Simons
gravity. We have determined quantitatively the difference between Kerr and CS predictions, and find
that this difference is well below detection limit for all observables, at the maximum level of $0.1\%$. 

Our first conclusion is that slow-rotation,
small-coupling CS theory will not be tested by electromagnetic observations in the vicinity of black holes,
at least in the foreseeable future.

Our second conclusion is that in order to determine whether electromagnetic signatures 
of black holes can help test CS gravity,
it is necessary to develop numerical solutions of CS black holes allowing arbitrarily high values
of spin. {Such solutions would still be at leading order in the coupling constant, as CS gravity is an effective 
theory (ghosts modes are likely to develop in an exact theory, see the discussion in~\citep{yagi12})}. 
It is most probable that higher order perturbative expansion of the solution, 
as was very recently proposed by Ref.~\citep{yagi12}, will not change our first conclusion. 
{The metric found by these authors give corrections to the diagonal metric coefficients
that can be as high as $1/r$. However, the relative difference $\delta g_{\mu\nu}^{\mathrm{CS,2}}/g_{\mu\nu}^{\mathrm{Kerr}}$
of the second-order CS corrections to the Kerr metric coefficients are always of order $1/r^3$ (just as for the ratio
$\delta g_{t\varphi}^{\mathrm{CS,1}}/g_{t\varphi}^{\mathrm{Kerr}}$ in the first-order solution). Given the extremely
small observable differences that we obtained in the first-order expansion, it is unlikely that a second-order
treatment would lead to a different conclusion. In order to obtain a more quantitative feeling of the effect of
using the second-order CS solution, we computed the relative difference of the radial epicyclic frequency profiles in the Kerr
solution and in the second-order CS solution, keeping only the leading-order term in all $g_{\mu\nu}$ 
coefficients given in Ref.~\citep{yagi12}. This relative difference is still of the order of a fraction of a percent,
just as in the first-order CS case.}
Numerical solutions of CS gravity were already derived for slowly rotating black holes, with arbitrary
value of the coupling constant~\citep{haimoud11}. 
We think that it is in this way that future constraints
may be obtained for electromagnetic observables.


\ack
FHV acknowledges fruitful discussions on CS and other alternative theories of gravitation 
with W. Klu\'zniak and M. Abramowicz.



\bibliographystyle{unsrt}

\end{document}